# AI PROGRESS IN SKIN LESION ANALYSIS

Philippe M. Burlina, PHD[1,2], William Paul[1], Phil A. Mathew[1], Neil J. Joshi, BS[1], Alison W. Rebman, MPH[3], John N. Aucott, MD[3]

[1]Applied Physics Laboratory, Johns Hopkins University

[2]Malone Center for Engineering in Healthcare, Johns Hopkins University

[3]Johns Hopkins Lyme Disease Research Center, Division of Rheumatology,

Department of Medicine, Johns Hopkins University School of Medicine

**ABSTRACT**

We examine progress in the use of AI for detecting skin lesions, with particular emphasis on the erythema migrans rash of acute Lyme disease, and other lesions, such as those from conditions like herpes zoster (shingles), tinea corporis, erythema multiforme, cellulitis, insect bites, or tick bites. We discuss important challenges for these applications, in particular the problems of *AI bias* regarding the lack of skin images in dark skinned individuals, being able to accurately *detect, delineate, and segment* lesions or regions of interest compared to normal skin in images, and *low shot learning* (addressing classification with a paucity of training images). Solving these problems ranges from being highly *desirable requirements* -- e.g. for delineation, which may be useful to disambiguate between similar types of lesions, and perform improved diagnostics -- or *required* – as is the case for AI de-biasing, to allow for the deployment of fair AI techniques in the clinic for skin lesion analysis. For the problem of low shot learning in particular, we report skin analysis algorithms that gracefully degrade and still perform well at low shots, when compared to baseline algorithms: when using a little as 10 training exemplars per class, the baseline DL algorithm performance significantly degrades, with accuracy of 56.41%, close to chance, whereas the best performing low shot algorithm yields an accuracy of 85.26%.

**I Introduction**

Lyme disease is the most common tick-borne disease in the United States, with an estimated 300,000 new cases per year.[1–3] The bacteria *Borrelia burgdorferi* is the causative vector of Lyme disease in North America. This bacterial agent is inoculated into humans through an infected tick bite. After 3 to 30 days, a round or oval, red, centrifugally expanding skin lesion (erythema migrans or EM) manifests itself in approximately 70-80% of cases.[4,5] EM may also be accompanied by flu-like symptoms including fever, fatigue, myalgia, and arthralgia. Without appropriate antibiotic treatment, EM can persist and subsequently resolve under pressure from the immune response.[4] Artificial intelligence (AI) can play a key role in the detection of EM, which is important for initiating early treatment for Lyme disease, without which the bacteria may disseminate into the nervous, cardiac, and rheumatologic systems.[6–19]

Deep learning (DL) has led to critical successes in tackling various AI tasks, including automated image classification for the purposes of medical diagnostics [20–48] such as in retinal analytics [22–26,48] and also, related to this work, cancerous skin lesions.[47] We have been developing AI tools for other types of skin analytics, specifically the diagnosis of non-cancerous skin lesions including erythema migrans (for Lyme disease), herpes zoster (shingles), tinea corporis, erythema multiforme, cellulitis, insect bites, or tick bites, using clinically obtained images as well as public domain data obtained from the internet [33,49]

In the current manuscript, we examine advances in the use of AI for detecting skin lesions and specifically the erythema migrans rash of acute Lyme disease. In particular, we report on recent developments that address important challenges for these applications, including the problem of AI bias, which may affect classification in individuals with darker skin; the problem of addressing a paucity of training skin image exemplars of lesions, so called 'low shot' learning; and the problem of automated segmentation and detection of skin lesions, which can play an important role in characterizing the progression of skin lesions, particularly EM, over time and can aid in the detection of Lyme disease.

## 2 Low Shot Learning of Skin Lesions

We have recently demonstrated that AI methods can be successfully used for the detection of EM and other skin lesions.[33,49] However, one key weakness of AI via DL is the need for very large ground truth-annotated training datasets, often requiring hundreds if not thousands of images. Training based on only a few (or a dozen) exemplars is called 'low shot' or 'few shot' learning and is a challenge which humans have the ability to perform innately well compared to machines: consider, for example, a child who is easily able to differentiate zebras from horses after being shown only one image of a zebra, thus performing one shot learning.

By contrast, 'traditional' AI algorithms generally lack the ability to adequately train with low numbers of gold standard training images, a challenge which remains an open and active problem of investigation. With only a modest number of training images available, traditional DL approaches often perform very poorly.[50] Datasets with only a few image training examples, under certain circumstances, may also yield low testing power from which reliable conclusions cannot be inferred.

There are several important, practical settings in which this low shot problem is relevant for skin lesion diagnostics or when using DL methods for automated skin analysis. One is in the case of rare diseases, for example erythema marginatum,[51] or more unusual presentations of EM such as those with vesiculation.[19]

We have worked towards utilizing novel methods in addressing low shot learning as it applies to skin diseases and in particular, Lyme disease. We have relied on techniques including the use of self-supervision, which does not require gold standard labels and thus can exploit data that do not have annotations.[52–54] In self-supervised methods, part of the image is used to predict another part of the same image, which results in a network that can be used for representation learning. An example of such a network is Deep InfoMax,[53] which is trained to maximize the similarity of local and global representations of images at different layers of the network, with the same source image as input, and minimize the similarity of pairs not created from the same input image. It uses an information-theoretic framework as a loss function.

We performed a comparison of different approaches towards low shot learning for skin analysis which included the following:

*Classical Baseline Deep Learning System.* We used a baseline method using a classical DL system, i.e. ResNet50, and performed fine-tuning. We denote it henceforth as "RES". This method has been widely used in medical AI diagnostic problems. This method was compared against novel systems which allow for *Low Shot Deep Learning* (LSDL), including:

*Low Shot Deep Learning Approaches via discriminative encoding.* We used methods relying on representation learning via encoding of images using a discriminative neural network. These methods used global pooling of ResNet50's[41,45] last convolutional layers. We then used one of three classifiers including support vector machines, random forest, and K-nearest neighbors, respectively abbreviated as RES_Random Forest, RES_RBF SVM and RES_KNN.

*Low Shot Deep Learning (LSDL) Approaches via self-supervision.* A second family of LSDL methods that were used here worked via self-supervision and proceeded also by first encoding images via a self-supervised network then using additional classification logic. We used a specific variant of the Deep InfoMax approach[54] that makes use of different image augmentations (color jitter, random resized crop of the same image), to compute the similarity between global and local representations of the image. As one of the LSDL algorithms, henceforth abbreviated as DIM, we used the local representation as input to train a ResNet network. Also, as in the previous group of methods, we also considered additional LSDL methods that instead used the global representation, which was then fed to three classifiers (support vector machines, random forest, and K-nearest neighbors, yielding methods which are henceforth referred to as DIM_Random Forest, DIM_RBF SVM, and DIM_KNN).

We tested and compared the performance of the above algorithms using a binary problem consisting of classifying Lyme vs healthy. We used a dataset of image provisioned online and curated by our team. [49] We show promising results in Table 1 that demonstrate that low shot learning techniques may be beneficial for EM detection, and demonstrate graceful degradation when compared to traditional DL methods. As seen in the table, while all methods perform about the same at high shots (N=5120), when using a little as N=10 training exemplars per class, the baseline algorithm performance (RES) significantly degrades, with accuracy of 56.41%, close to chance, whereas the best performing low shot algorithm (DIM_RBF SVM) yields an accuracy of 83.33% and 85.26% for RES_Random Forest.

| Number of Shots | DIM | DIM_KNN | DIM_Random Forest | DIM_RBF SVM | RES | RES_KNN | RES_Random Forest | RES_RBF SVM |
| --- | --- | --- | --- | --- | --- | --- | --- | --- |
| 5120 | 91.67 | 84.62 | 88.46 | 91.03 | 94.23 | 87.82 | 89.1 | 89.74 |
| 639 | 91.67 | 84.62 | 87.82 | 91.03 | 95.51 | 87.82 | 89.74 | 89.74 |
| 79 | 88.46 | 82.05 | 82.69 | 87.18 | 81.41 | 83.97 | 87.18 | 88.46 |
| 40 | 81.41 | 82.69 | 83.97 | 88.46 | 74.36 | 83.33 | 85.9 | 87.18 |
| 20 | 76.92 | 75.64 | 83.97 | 82.69 | 48.72 | 83.33 | 83.97 | 87.82 |
| 10 | 79.49 | 78.85 | 78.21 | 83.97 | 56.41 | 77.56 | 85.26 | 83.33 |

**Table 1** this table illustrates the benefit of each low shot methods' performance (as measured via accuracy, in percentage) when compared to the baseline method (RES), when the number of training exemplars per class (the number of shots N) decreases from 5120 down to only 10 exemplars. The method RES exemplifies a traditional fine-tuned ResNet50 method and shows significant degradation for N=20 and N=10, whereas other methods have a more graceful degradation and better performance even for those low values of N (e.g. DIM_RBF SVM).

## 3 Automated Segmentation of Skin Lesions

Skin lesion segmentation is an important task as it would allow for time-based analysis of the evolution of the lesion. This task may help in diagnosing lesions, particularly those where there may often be a degree of ambiguity, such as in the case of EM. Unlike traditional approaches in computer vision using graph based methods for segmentation[55] we are pursuing here automated delineation through the use of fully convolutional networks, [56,57] Examples of results obtained using this technique for skin delineation are shown in Figure 1. Our work is currently directed at addressing some challenges in EM skin segmentation when dealing with poorly defined lesion boundaries or lesions that do not appear as the classic "bullseye" shape.

As an alternate task to specific tissue delineation, one task which is useful to aid in skin analysis and skin lesion diagnosis is the detection of the areas of interest on the skin using object detectors. Examples of this technique are demonstrated in Figure 2, where we aimed to find areas of redness around osseointegration sites and external pins in patients with prosthetics or external fixators used to stabilize broken bones, where a redness could suggest the presence of an infection which may need to be treated.

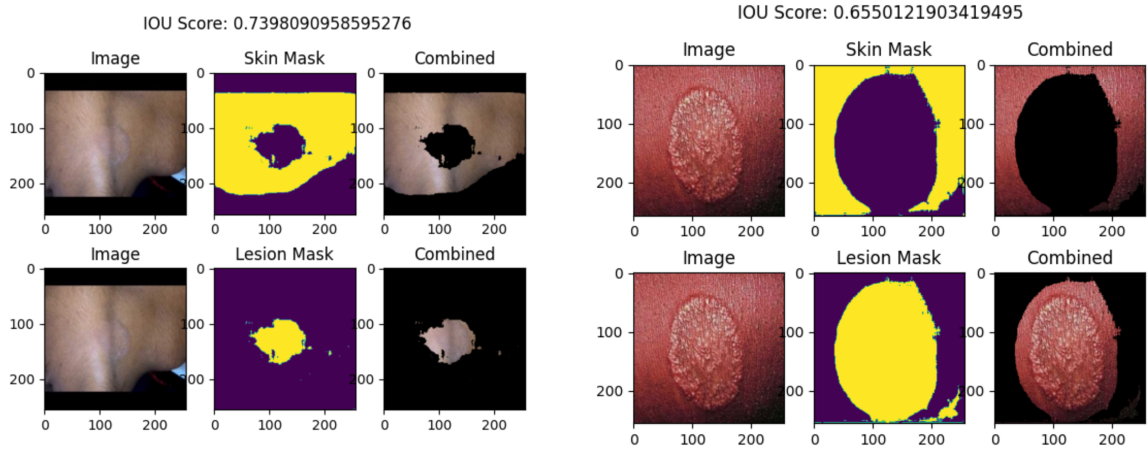

Figure 1: Examples of segmentation of skin (top) and lesion (bottom) applied here to tinea corporis.

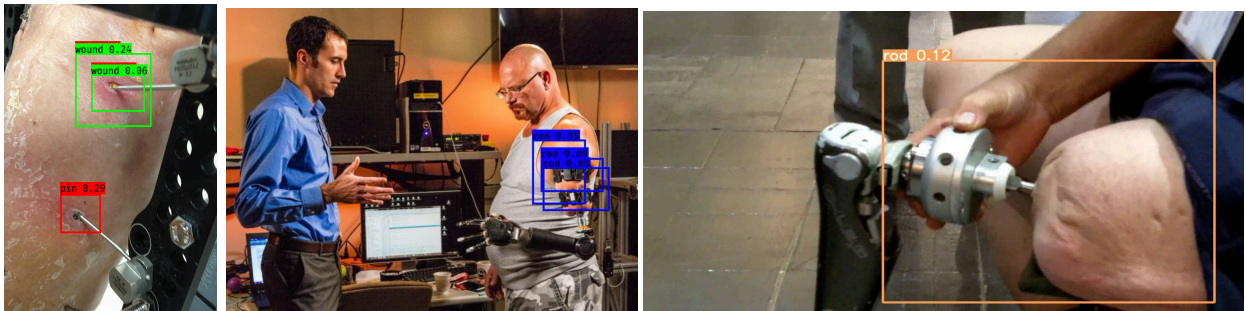

Figure 2: Example of applying object detectors for finding lesions and areas of interest in skin images such as fixator pins or prosthetics' osseointegration sites.

**3 Bias in AI**

The problem of AI bias with regard to factors such as age, gender, or skin color has gained significant attention recently as one of the key AI assurance issues that will need to be addressed for fielding of AI systems. Of the several potential causes of AI bias, one could occur when the majority of data collected and used in training AI algorithms is unbalanced and specific demographic groups are overrepresented.[58] In our prior studies on automated tools that seek to find online images of skin lesions,[33,49] we have found that an overarching amount of data available online and in the public domain includes images almost exclusively from lighter skinned individuals, and many fewer images of darker skinned individuals. Studies have also shown that medical textbooks on dermatological lesions have similar imbalances in their teaching images, with most representative of lighter skinned individuals. Consequently, 47% of dermatologists report a lack of training or exposure to patients with darker skin.[59]

Most current AI algorithms are data-driven. Therefore, classification performance of these algorithms depends on available data from large cohorts of individuals where a gold standard diagnosis was determined for each individual[58] and good balance exists between subcategories and subpopulations. Often, the root cause of the AI bias problem may be a lack of balance in the AI training datasets, that is, the dataset might have been from predominantly light-skinned individuals. The problem may also have other causes than just data. For example, the data may be balanced but human annotators may be biased, or the quality and diversity of the data may vary by specific demographic groups, or the algorithm may use indicators for performing predictions (e.g. for predicting the need for healthcare) that may perpetuate bias. This problem can sometimes also be understood to be a domain generalization problem of ML algorithms. Often finding the root cause of biased AI can be a challenging problem in and of itself.

It turns out that there are many different definitions of bias as well as objectives regarding de-biasing. The type of diagnosis performance problem occurring when AI performance varies depending on a demographic subpopulation is referred to as "(in)equality of odds" and the desired behavior is that protected factors like gender, race, or age of the individual should not affect diagnosis performance. Other issues have been widely reported in other domains, such as for example the COMPAS system (correctional offender management profiling for alternative sanctions) that assesses recidivism risks, which has been shown to predict risk differentially by race, even when such differences did not exist. This problem may be referred to as '(un)equal opportunity' type of bias.

We have developed AI tools[60] that help alleviate AI bias in retinal diagnosis in patients with light vs. darker skin and found different outcomes, with lower performance among darker skin individuals. We also found that algorithms which were developed using data from mostly white individuals had lower performance when tested on patients from Asian ethnic groups. In other recent experiments working with mental health data we also found that unbalanced data can lead to bias.

One of our goals is to study the problem of AI bias for skin diseases. In a preliminary study, we found that for certain types of skin lesions, and for simpler problems, a lack of balance in the data does not necessarily lead to bias (Figure 3), It may lead to bias in more complex ones. In the future, we intend to collect data from darker skinned individuals with EM, which may be challenging, and investigate the sensitivity of classification results to the lack of balance in the data. Potential strategies for the future collection of skin lesion images from darker skinned individuals include collaborating with dermatologists who routinely care for individuals with darker skin, and continuing to mine publicly available or other research datasets for relevant images.

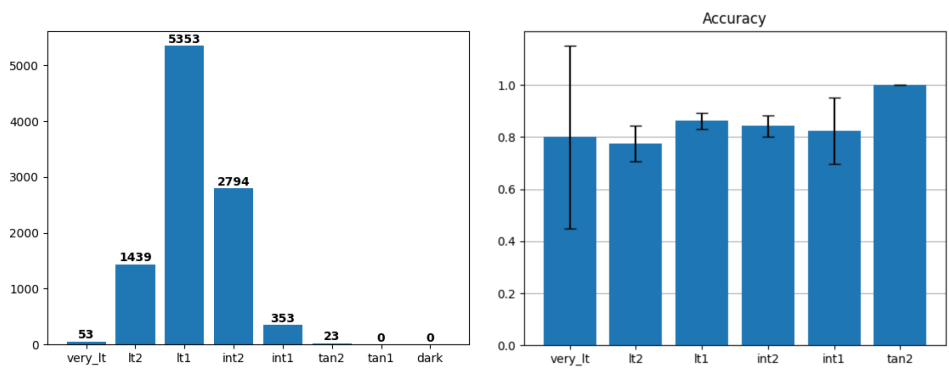

Figure 3 A study of bias from ISIC2018 dataset of dermoscopic imagery and skin cancer lesions: while imbalance may exist among populations with varying skin tones going from very light (marked as 'very_lt') to darker (eg 'tan2') (see also this work[37] which first reported this finding), this does not always lead to AI bias in this relatively benign use case. We will probe further such issues of bias in images taken 'in the wild' (arbitrary background, view angles and illumination) of more complex lesions and confusers including EM and tinea corporis.

## 5 Discussion and Conclusion

We have reported on several important challenges in the use of AI for detection of erythema migrans and other skin disease. We have shown that it is possible to address low shot learning and training with much less exemplars. We have also reported preliminary results for skin lesion delineation and detection. Doing skin lesion delineation can enhance the performance of diagnostic algorithms by allowing the assessment of the time evolution of the lesion. This time evolution can be further characterized using a combination of automatic segmentation techniques and tracking algorithms such as[61], a possibility we will explore in the future.

We have started investigating the problem of AI bias for skin lesions. Testing for the presence or absence of bias is the first important step that should be taken prior to fielding any AI application in the clinic, as often the harm is not recognizing that there is even an issue of bias in the deployment of such AI algorithms.

Finding the cause of the bias is the next step, and it is not always an easy task. Data (information bias) are not the only source of bias. Other sources include imbalances in the individuals selected for the data (selection bias) and in the techniques used to obtain those data (ascertainment bias). Assumptions which the algorithm makes can also be a source of bias, and there can be many other additional causes. Sometimes the challenge is not necessarily a lack of data from specific demographic (age, gender, or racial/ethnic) subgroups but rather recognition that the data used to train a neural network may yield a model that does not always generalize to the subsequent data being evaluated by that neural network. In some cases, it is possible to acquire more data, and in the case of prospective data collection, experiments should be designed so that demographically-defined subgroups are appropriately represented. In some cases when using retrospective data, one may not have the option to acquire more data in order to reduce bias. This is where specific AI techniques may come in to allow some form of intelligent data augmentation to address the lack of data. Those types of AI techniques, called generative models, are able to generate new data that may be missing or to take a dataset that is biased and generate a new, unbiased dataset from which machines can then learn to predict equitable outcomes. For other areas like retinal analysis, we have shown that in some cases this strategy can help address bias issues in AI algorithms. In future analyses, we plan to apply and investigate such techniques for skin analytics. Furthermore, there are other approaches that we envision for future studies that are not data-based but more algorithmic-based. We will also be looking at developing techniques that may help find the root cause and sources of bias. We hope that by using such techniques, we can also help train humans with balanced data and address some bias that may exist in clinicians.